\documentclass{AIAA}

\begin{document}

\title{Relativistic Dynamics of Relative Motions (I): Post-Newtonian
Extension of the Hill-Clohessy-Wiltshire Equations}

\author{Peng Xu\footnote{Research Associate, Institute of Applied Mathematics, xupeng@amss.ac.cn}}
\affiliation{Academy of Mathematics and Systems Science, Chinese Academy of Sciences,
Beijing, 100190, China}

\author{Li-E Qiang\footnote{Lecturer, Department of Geophysics, College of the Geology Engineering and
Geomatics, qqllee815@chd.edu.cn}}
\affiliation{Chang'an University, Xi'an, 710054, China}

\begin{abstract}
With continuous advances in technologies related to deep space ranging and satellite gravity gradiometry, corrections
from general relativity to the dynamics of relative orbital motions will certainly become important. In this work, we extend,
in a systematic way, the Hill-Clohessy-Wiltshire Equations to include the complete first order post-Newtonian effects
from general relativity. Within certain short time limit, post-Newtonian corrections to general periodic solutions of the
Hill-Clohessy-Wiltshire Equations are also worked out.
\end{abstract}

\maketitle

%\section*{Nomenclature}
%\noindent\begin{tabular}{@{}lcl@{}}
%$a$   &=&    semimajor axis \\
%$c$  &=& speed of light\\
%$e_{(a)}{}^{\mu}$ &=& tetrad matrix\\
%$i$ &=& inclination\\
%$g_{\mu\nu}$ &=& spacetime metric\\
%$G$ &=& gravitational constant\\
%$M$ &=& total mass of Earth \\
%$\vec{J}$ &=& angular momentum of Earth\\
%$J_2$ &=& equatorial bulge gravitational potential coefficient\\
%$x^{\mu}=\{t,\ x^i\}$ &=& coordinates of geocentric system\\
%$\omega$ &=& angular grequency\\
%$\Omega$ &=& longitude of ascending note\\
%$\Psi$ &=& true anomaly\\
%$\tau$ &=& proper time measured along the orbit\\
%$\Gamma^{\mu}_{\ \ \lambda\rho}$ &=& Christoffel Symbol\\
%\textit{Subscript}\\
%$i,\ j,\ k,...$ &=& indices of spatial tensor components \\
%$\mu,\ \nu,\ \rho,\ \lambda,...$ &=& indices of spacetime tensor components
%\end{tabular}

\section{Introduction and Motivations\label{sec:Introduction}}

The studies of relative motions between orbiting satellites can date back
to early 1960s, which, by then, was motivated mainly by the satellites rendezvous
problem. Clohessy and Wiltshire in their celebrated work \cite{Clohessy1960}
analyzed the motion of a satellite with respect to a reference one
that following a circular orbit around an uniform and spherical
gravitational source. With the assumption that the distance between
the two satellites is short compared with the orbital radius, the system of linear
equations around the null solutions, that the so-called Clohessy-Wiltshire equations, were derived.
Such equations are also known as Hill's equations, since
Hill had employed the same method to study the motions of the Moon with respect to
the Sun-Earth system \cite{Hill1878}.
The Hill-Clohessy-Wiltshire (HCW)
equations and the related methods had then found wide applications.
In-depth analysis including nonlinear effects, eccentricity,
central source oblateness, drags, and other perturbations had also been
carried out.
%Due to the progress of satellites formation flight in the past decays, the studies of
%relative motion theories became rather active with many different approaches been developed \cite{Gim2003,Alfriend2002,Sengupta2003}.

The HCW equations had proved useful to satellite geodesy,
such as to the error analysis of satellite altimetry \cite{Schrama1989},
computations of the ephemerides of GPS orbits \cite{Colombo1989}, and the analysis
of low-flying Earth orbiters \cite{Schrama1991,Sneeuw1994}. In
the pioneering work of Wolff \cite{Wolff1969}, it was pointed out that the
global variations and intensities of the Earth geopotentials could be
mapped out given the range and range rate data of a Satellite-to-Satellite
Tracking (SST) system. Based on such principle, the Gravity Recovery
and Climate Experiment (GRACE) mission %\footnote{{http://www.csr.utexas.edu/grace/} and
%{http://www.jpl.nasa.gov/missions/current/grace.html}}
had continuously provided us valuable
data through its microwave inter-satellites range measurements in the past 14 years.
To continue the measurements of the Earth geopotential variations, NASA had scheduled the
launch of the GRACE Follow On mission, which will be supplemented
with a laser interferometer of nanometer-level accuracy. %\footnote{{http://gracefo.jpl.nasa.gov/}}.
GRACE II %\footnote{{http://decadal.gsfc.nasa.gov/GRACE2.html}}
and
other possible Next Generation Gravity Missions (NGGM) are also
under investigations. The HCW equations had played
an important part in the pre-mission analysis and the gravity field
determinations of such SST missions \cite{Colombo1984,Colombo1986,Mackenzie1997}. Based on the HCW equations, the so called
semi-analytic method of gravity field analysis from satellite
observations had also been developed \cite{Sneeuw2000}.
%Furthermore, the HCW equations with higher
%order corrections was applied to the orbit designs [?]
%of the Laser Interferometer Space Antenna of gravitational waves
%(eLISA, classed as ESA's L3 mission and scheduled in 2034) \footnote{{https://www.elisascience.org/}}.
With the continuous advances in  technologies related to deep space ranging, especially
the  laser interferometry in space, the
sensitivities and resolutions of NGGMs that based on measurements of
inter-satellites ranges or distances between orbiting proof masses will
be greatly improved. Therefore, general relativistic corrections from spacetime curvature to
the dynamics of relative orbital motions will certainly become important and can not be ignored.

As an example, the LISA PathFinder (LPF) mission \cite{Armano2016}
 carried two proof masses that were successfully put in pure gravitational
free fall with acceleration noise maintained to $5.2\pm0.1\ fm/s^{2}\sqrt{Hz}$
in the frequency band $0.7\ mHz\sim20\ mHz$, and its on-board laser interferometer
as the readout of the relative motions between the proof masses had
the sensitivity better than $9\ pm/\sqrt{Hz}$ in the same band.
The measurement scheme of the LPF can be viewed as a
demonstration of an one-dimensional optical gravity gradiometer, and its success may
pave the way of applying high-precision optical gradiometers in future gravity
missions (%see the projects A07 and B07 of
e.g. the geo-Q mission).
%\footnote{{http://www.geoq.uni-hannover.de/a07.html} and {http://www.geoq.uni-hannover.de/b07.html}}).
With such high sensitivities, the advantage of using orbiting optical gradiometer in relativistic experiments
are now under investigations \cite{Xu2016}.
Also, for SST missions like the GRACE, it is known that certain effects from
General Relativity (GR) need to be considered (numerically)
in data analysis. And, it had been noticed that tests of relativistic gravitational theories
may be carried out with the observations from the GRACE Follow On \cite{Qiang2015} and future
satellite gravity gradiometry missions \cite{Mashhoon1989,Qiang2015b,Qiang2016}.
Therefore, to provide the theoretical tools, it is meaningful to generalize, in a systematic and self-consistence way,
the HCW analysis of relative motions to include corrections from GR.

In the next section, a systematic approach, through the (Jacobi) geodesic deviation
 equation, of analyzing the linearized dynamics of relative
motions between orbiting satellites or proof masses in GR is explained
in a self-contained manner. With models and notations introduced, from Sec. \ref{sec:Reference orbit} to \ref{sec:PNHCW},
we derive the generalized HCW equations including the complete first order Post-Newtonian (PN)
\cite{Will1993,Will2014} effects from GR under the conditions of weak fields and slow motions in Solar system.
Within certain short time limit, the general PN corrections to periodic
solutions of the classical HCW equations are worked out analytically.

\section{Linearized Dynamics of Relative Motions in GR\label{sec:Linearized in GR}}

Einstein's general theory of relativity is a geometric theory of gravitation. In the past few decays, the
fundamental principle, that the Einstein's equivalence principle, behind such geometric theory
and the many critical predictions drawn from GR (including gravitational waves
\cite{Abbott2016}) had been well-tested with great accuracies \cite{Turyshev2008,Will2014}. Today,
GR is still the best fit relativistic theory of gravitation among
the many alternatives \cite{Will2014}.

In GR, satellites or proof masses freely falling in gravitational
field will follow geodesic world lines in a 4-dimensional spacetime,
which extremize the action (length of world lines)
\begin{equation}
S=\int\sqrt{-g_{\mu\nu}\tau^{\mu}\tau^{\nu}}d\tau\label{eq:action}
\end{equation}
where $g_{\mu\nu}$ is the spacetime metric field, $\tau^{\mu}=dx^{\mu}/d\tau$
the 4-velocity of the satellite or proof mass and $\tau$ the proper
time measured along the world line. In this work, we use $i,j,k,...=1,2,3$
to index the spatial tensor components and $\mu,\nu,\lambda,...=0,1,2,3$
the spacetime tensor components, and the Einstein summation convention
is assumed. For Solar system experiments, the typical speed $v$ of orbiting
satellites is much smaller compared with the speed of light $c$, especially
for low or medium Earth orbits the ratio $v/c=\epsilon$ is about $10^{-5}\sim10^{-6}$.
Also, according to the Virial theorem for Newtonian system, one has
$v^{2}/c^{2}\sim U/c^{2}\sim\mathcal{O}(\epsilon^{2}),$ where $U$
is the Newtonian potential and the dimensionless quantity $U/c^{2}$
is a measure of the strength of the gravitational field. Therefore,
Newtonian gravity can be viewed as the weak field and slow motion
approximation of GR up to $\mathcal{O}(\epsilon^{2})$. For satellite
gravity missions and space-borne relativistic experiments in the present-day
and near future, it is sufficient to keep the (dimensionless) working
precision up to $\mathcal{O}(\epsilon^{4})$, which just gives rise to the
Post-Newtonian approximation of GR \cite{Will1993,Will2014}. For clarity,
the geometric units $G=c=1$ are adopted hereafter, and
in Sec. \ref{sec:PNHCW} the SI units will be recovered. The action in Eq.(\ref{eq:action})
may be expanded as
\begin{eqnarray}
S & = & \int\sqrt{-g_{00}(\frac{dx^{0}}{d\tau})^{2}-2g_{0i}\frac{dx^{0}}{d\tau}\frac{dx^{i}}{d\tau}-g_{ij}\frac{dx^{i}}{d\tau}\frac{dx^{j}}{d\tau}}d\tau\nonumber \\
 & \sim & \int\sqrt{-g_{00}-2g_{0i}v^{i}-g_{ij}v^{i}v^{j}}d\tau\nonumber \\
 & \sim & \int\sqrt{-g_{00}-2g_{0i}\mathcal{O}(\epsilon)-g_{ij}\mathcal{O}(\epsilon^{2})}d\tau.\label{eq:action_order}
\end{eqnarray}
Therefore, to study geodesic or orbit motions up to the PN level the
metric components have to be expended to
\begin{equation}
g_{00}\sim-1+\mathcal{O}(\epsilon^{4}),\ \ g_{0i}\sim\mathcal{O}(\epsilon^{3}),\ \ g_{ij}\sim1+\mathcal{O}(\epsilon^{2}).\label{eq:metric_order}
\end{equation}

In this work, we model Earth as an ideal uniform
and rotating spherical body, with total mass $M$ and angular momentum
$\vec{J}$. The inertia and geocentric Cartesian coordinates system $\{t,\ x^{i}\}$
is chosen that one of its spatial basis $\frac{\partial}{\partial x^{3}}$
is parallel to the direction of $\vec{J}$ and the coordinate time
$t$ is measured by observers in the asymptotically flat region.
According to Eq. (\ref{eq:metric_order}), the PN metric
outside Earth may be expanded as \cite{Weinberg1972,Straumann1984}
\begin{equation}
g_{\mu\nu}=\left(\begin{array}{cccc}
-1+2U-2U^{2} & \frac{2x^{2}J}{r^{3}} & -\frac{2x^{1}J}{r^{3}} & 0\\
\frac{2x^{2}J}{r^{3}} & 1+2U & 0 & 0\\
-\frac{2x^{1}J}{r^{3}} & 0 & 1+2U & 0\\
0 & 0 & 0 & 1+2U
\end{array}\right),\label{eq:metric}
\end{equation}
where $r=\sqrt{(x^{1})^{2}+(x^{2})^{2}+(x^{3})^{2}}$. The
PN order relations for an orbiting satellite or proof mass read
\begin{eqnarray}
v^{2} & \sim & \frac{M}{r}\sim\mathcal{O}(\epsilon^{2}),\label{eq:MJorder1}\\
v^{4} & \sim & \frac{M^{2}}{r^{2}}\sim\frac{Jv}{r^{2}}\sim\mathcal{O}(\epsilon^{4}).\label{eq:MJorder2}
\end{eqnarray}
Deviations from uniform sphere of the centered gravitational source will give rise to
corrections to the above metric, and their
main contributions will be the geopotential multipoles (in terms of
the spherical harmonics) added to the Newtonian potential $U$ in the time-time
component of the metric
\begin{equation}
g_{00}=-1+2\frac{M}{r}(1+\sum_{l=2}^{\infty}(\frac{R}{r})^{l}\sum_{m=-l}^{l}C_{lm}Y_{lm})-\frac{2M^{2}}{r^{2}},\label{eq:g00+multiples}
\end{equation}
where $R$ is the averaged radius of Earth. While, since $J_{2}$ is a rather
large component, which is only about $4\times10^{-4}$ times smaller than
the monopole field $\frac{M}{r}$ of Earth. Therefore, considering
the possible sensitivities and resolutions for future gravity missions,
the relativistic corrections from $J_{2}$ should also be included
and the more accurate metric turns out to be{\footnotesize{}
\begin{eqnarray}
 &  & g_{\mu\nu}=\nonumber \\
 &  & \left(\begin{array}{cccc}
-1+\frac{2M}{r}(1+\sum_{l=2}^{\infty}(\frac{R}{r})^{l}\sum_{m=-l}^{l}C_{lm}Y_{lm}) & \frac{2x^{2}J}{r^{3}} & -\frac{2x^{1}J}{r^{3}} & 0\\
-\frac{2M^{2}}{r^{2}}-\frac{4C_{20}R^{2}M^{2}}{r^{4}}Y_{20}\\
\\
\frac{2x^{2}J}{r^{3}} & 1+2(\frac{M}{r}+\frac{C_{20}R^{2}M}{r^{3}}Y_{20}) & 0 & 0\\
\\
-\frac{2x^{1}J}{r^{3}} & 0 & 1+2(\frac{M}{r}+\frac{C_{20}R^{2}M}{r^{3}}Y_{20}) & 0\\
\\
0 & 0 & 0 & 1+2(\frac{M}{r}+\frac{C_{20}R^{2}M}{r^{3}}Y_{20})
\end{array}\right).\nonumber \\
\label{eq:full_metric}
\end{eqnarray}
}In this first-step theoretical study of the relativistic dynamics of relative
motions, we will restrict ourselves to work with the metric
field of the ideal Earth model given in Eq. (\ref{eq:metric}).
Subsequent studies including perturbations from certain multipoles based on the metric in Eq. (\ref{eq:full_metric})
and the possible applications to SST missions like the GRACE and GRACE Follow On, especially the effects on the measurement accuracy of the $J_2$ component, will be left in a separated publication.

Given the metric field, the relative motions among a family of adjacent free-falling
proof masses or satellites are driven by the spacetime curvature. Especially, when the distance between the adjacent
two satellites or proof masses is smaller compared with the curvature
radius of the corresponding spacetime region, their relative motion
satisfies the so called (Jacobi) geodesic deviation equation
\cite{Wald1984}
\begin{equation}
\tau^{\rho}\nabla_{\rho}\tau^{\lambda}\nabla_{\lambda}Z^{\mu}+R_{\rho\nu\lambda}^{\ \ \ \ \mu}\tau^{\rho}\tau^{\lambda}Z^{\nu}=0,\label{eq:deviation}
\end{equation}
which is evaluated along the world line of the reference satellite or mass.
Here $\nabla_{\mu}$ denotes the covariant derivative associate to
the given metric, $R_{\rho\nu\lambda}^{\ \ \ \ \mu}$ the Riemann
curvature tensor, and $Z^{\mu}$ is the position difference 4-vector
(connection vector) pointing from the reference satellite or mass to the second one. This
is a linear equation of the position difference $Z^{\mu}$,
and effects from spacetime curvature can be interpreted as tidal forces
under local inertia frames (Fermi-shifted local frame \cite{Wald1984}) carried by the reference satellite or mass,
see \cite{Ni1978} for details.

\begin{figure}
\includegraphics[scale=1.2]{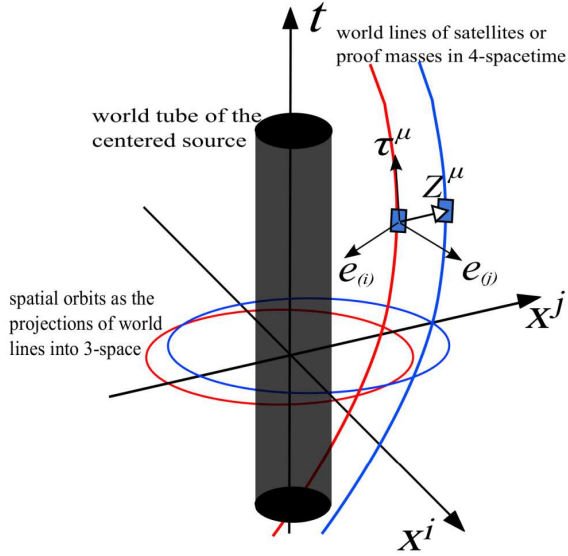}
\caption{The illustration of the world lines of adjacent orbiting satellites or proof masses in spacetime.
The spatial orbits can be viewed as the projections of the world lines into the 3-space.
The variations of the connection vector $Z^{\mu}$, or the relative motions between adjacent satellites or masses,
 is determined by spacetime curvature.}\label{fig:WL}
\end{figure}

Back to our problem, the world lines in
spacetime corresponding to the adjacent orbits of
satellites or proof masses can be illustrated in Fig. \ref{fig:WL}. In the local frame, which is defined by
the tetrad  $e_{(a)}^{\ \ \ \mu}$ ($a=0,1,2,3$) attached to the reference satellite or mass with $e_{(0)}^{\ \ \ \mu}=\tau^{\mu}$,
the above geodesic deviation equation can be expanded as
\begin{eqnarray}
\frac{d^{2}}{d\tau^{2}}Z^{(a)} & = & -2\gamma_{\ \ (b)(0)}^{(a)}\frac{d}{d\tau}Z^{(b)}\nonumber \\
 &  & -(\frac{d}{d\tau}\gamma_{\ \ (b)(0)}^{(a)}+\gamma_{\ \ (b)(0)}^{(c)}\gamma_{\ \ (c)(0)}^{(a)})Z^{(b)}\nonumber \\
 &  & -K_{(b)}^{\ (a)}Z^{(b)},\label{eq:localdev}
\end{eqnarray}
where $Z^{(a)}e_{(a)}^{\ \ \ \mu}=Z^{\mu}$ and $\gamma_{\ \ (b)(c)}^{(a)}=e^{(a)\nu}\nabla_{\mu}e_{(b)\nu}e_{(c)}^{\ \ \ \mu}$
are the Ricci rotation coefficients \cite{Wald1984}. Here $\{(a),\ (b),...\}$ and $\{(i),\ (j),...\}$ are
used to index tensor components under the local frame.
The first line of the right hand side of the above equation contains the relativistic analogue of the
Coriolis force, the second line contains the inertia forces and the
third line is the tidal force from the spacetime curvature. The tidal
matrix is defined by
\begin{eqnarray}
K_{\mu\nu}& = & R_{\rho\mu\lambda\nu}\tau^{\rho}\tau^{\lambda},\label{eq:Kdefinition}\\
K_{(a)(b)} & = & K_{\mu\nu}e_{(a)}^{\ \ \ \mu}e_{(b)}^{\ \ \ \nu}.\label{eq:Klocal}
\end{eqnarray}

To summarize, Eq. (\ref{eq:localdev}) is the natural starting point
to analyze the linearized dynamics of relative orbit motions in GR.
As one will see, if all the relativistic effects are left out, Eq. (\ref{eq:localdev}) will reduce to the system of linear
equations of relative motions in Newtonian gravity.
In the following sections, we will evaluate Eq. (\ref{eq:localdev}) under the
local tetrad $e_{(a)}^{\ \ \ \mu}$
along a relativistic precessing circular orbit.
It is natural to choose the spatial bases $e_{(i)}^{\ \ \ \mu}$
to be the triad of the Local-Vertical-Local-Horizontal (LVLH) frame with relativistic corrections.
The explicit form of the PN extension of the HCW equations is derived, and the PN corrections to the periodic solutions of the HCW equations are worked out.

\section{Post-Newtonian Reference Orbit\label{sec:Reference orbit}}

\begin{figure}
\includegraphics[scale=0.9]{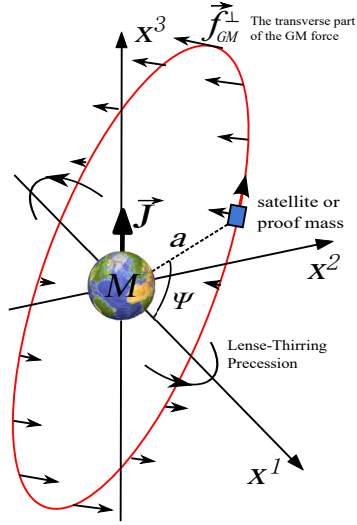}
\caption{The transverse part of the GM force $\vec{f}_{GM}^{\perp}$ acting on the satellite or proof mass
along a circular orbit and the Lense-Thirring precession of the orbit plane about the direction of $\vec{J}$.}\label{fig:orbit}
\end{figure}

As discussed in the last section, satellites or proof masses that
moving freely in gravitational field will extremize the action given
in Eq. (\ref{eq:action}), and therefore satisfy the following geodesic equation \cite{Weinberg1972,Wald1984}
\begin{equation}
\frac{d^{2}x^{\mu}}{d\tau^{2}}+\Gamma_{\:\:\rho\lambda}^{\mu}\frac{dx^{\rho}}{d\tau}\frac{dx^{\lambda}}{d\tau}=0,\label{eq:geodesic_tau}
\end{equation}
where $\Gamma_{\:\:\rho\lambda}^{\mu}$ denotes the Christoffel symbols.
Replacing the proper time $\tau$ with the coordinate time $t$ of
the geocentric coordinates system, the geodesic
equation becomes
\begin{eqnarray}
\frac{d^{2}x^{i}}{dt^{2}} & = & -\Gamma_{\:\:00}^{i}-2\Gamma_{\:\:0j}^{i}\frac{dx^{j}}{dt}-\Gamma_{\:\:jk}^{i}\frac{dx^{j}}{dt}\frac{dx^{k}}{dt}\nonumber \\
 &  & +(\Gamma_{\:\:00}^{0}+2\Gamma_{\:\:0j}^{0}\frac{dx^{j}}{dt}+\Gamma_{\:\:jk}^{0}\frac{dx^{j}}{dt}\frac{dx^{k}}{dt})\frac{dx^{i}}{dt}.\label{eq:geodesic_t}
\end{eqnarray}
Given the PN metric in Eq. (\ref{eq:metric}), the components of $\Gamma_{\:\:\rho\lambda}^{\mu}$ under the geocentric coordinates
system can be worked out up to the required order as
\begin{equation}
\Gamma_{\ \ 0\mu}^{0}=\frac{M}{r^{3}}\left(\begin{array}{c}
0\\
x^{1}\\
x^{2}\\
x^{3}
\end{array}\right),\label{eq:G1}
\end{equation}
\begin{eqnarray}
 \Gamma_{\ \ 0j}^{i}=
 \left(\begin{array}{ccc}
0 & -\frac{J\left((x^{1})^{2}+(x^{2})^{2}-2(x^{3})^{2}\right)}{r^{5}} & -\frac{3Jx^{2}x^{3}}{r^{5}}\\
\\
\frac{J\left((x^{1})^{2}+(x^{2})^{2}-2(x^{3})^{2}\right)}{r^{5}} & 0 & \frac{3Jx^{1}x^{3}}{r^{5}}\\
\\
\frac{3Jx^{2}x^{3}}{r^{5}} & -\frac{3Jx^{1}x^{3}}{r^{5}} & 0
\end{array}\right),\label{eq:G2}
\end{eqnarray}
\begin{equation}
\Gamma_{\ \ ij}^{1}=-\frac{M}{r^{3}}\left(\begin{array}{ccc}
x^{1} & x^{2} & x^{3}\\
x^{2} & -x^{1} & 0\\
x^{3} & 0 & -x^{1}
\end{array}\right),\label{eq:G3}
\end{equation}
\begin{equation}
\Gamma_{\ \ ij}^{2}=-\frac{M}{r^{3}}\left(\begin{array}{ccc}
-x^{2} & x^{1} & 0\\
x^{1} & x^{2} & x^{3}\\
0 & x^{3} & -x^{2}
\end{array}\right),\label{eq:G4}
\end{equation}
\begin{equation}
\Gamma_{\ \ ij}^{3}=-\frac{M}{r^{3}}\left(\begin{array}{ccc}
-x^{3} & 0 & x^{1}\\
0 & -x^{3} & x^{2}\\
x^{1} & x^{2} & x^{3}
\end{array}\right),\label{eq:G5}
\end{equation}
\begin{eqnarray}
 &  & \Gamma_{\ \ ij}^{0}=\nonumber \\
 &  & \left(\begin{array}{ccc}
\frac{6J\left(x^{2}(x^{1})^{3}+x^{2}\left((x^{2})^{2}+(x^{3})^{2}\right)x^{1}\right)}{r^{7}} & \frac{3J\left(-(x^{3})^{2}(x^{1})^{2}-(x^{1})^{4}+(x^{2})^{2}\left((x^{2})^{2}+(x^{3})^{2}\right)\right)}{r^{7}} & \frac{3J\left(x^{2}x^{3}(x^{1})^{2}+x^{2}x^{3}\left((x^{2})^{2}+(x^{3})^{2}\right)\right)}{r^{7}}\\
\\
\frac{3J\left(-(x^{3})^{2}(x^{1})^{2}-(x^{1})^{4}+(x^{2})^{2}\left((x^{2})^{2}+(x^{3})^{2}\right)\right)}{r^{7}} & -\frac{6J\left(x^{2}(x^{1})^{3}+x^{2}\left((x^{2})^{2}+(x^{3})^{2}\right)x^{1}\right)}{r^{7}} & -\frac{3J\left(x^{3}(x^{1})^{3}+x^{3}\left((x^{2})^{2}+(x^{3})^{2}\right)x^{1}\right)}{r^{7}}\\
\\
\frac{3J\left(x^{2}x^{3}(x^{1})^{2}+x^{2}x^{3}\left((x^{2})^{2}+(x^{3})^{2}\right)\right)}{r^{7}} & -\frac{3J\left(x^{3}(x^{1})^{3}+x^{3}\left((x^{2})^{2}+(x^{3})^{2}\right)x^{1}\right)}{r^{7}} & 0
\end{array}\right).\nonumber \\
\label{eq:G6}
\end{eqnarray}
Substitute $\Gamma^{\mu}_{\:\:\rho\lambda}$ into Eq. (\ref{eq:geodesic_t}) we then
have the equation of motion in PN approximations
\begin{equation}
\frac{d^{2}\vec{x}}{dt^{2}}=\vec{f}_{N}+\vec{f}_{GE}+\vec{f}_{GM},\label{eq:geodesic_Newton}
\end{equation}
where $\vec{f}_{N}=-\frac{M}{r^{3}}\vec{x}$ is the Newtonian force
per unit mass, and the relativistic corrections may be divided into
two parts that the GravitoElectric (GE) force and GravitoMagnetic
(GM) force per unit mass
\begin{eqnarray}
\vec{f}_{GE} & = & \frac{M}{r^{3}}\left((\frac{4M}{r}-v^{2})\vec{x}+4(\vec{x}\cdot\vec{v})\vec{v}\right),\label{eq:FGE}\\
\vec{f}_{GM} & = & 2\vec{v}\times\left(\frac{\vec{J}}{r^{3}}-\frac{3(\vec{J}\cdot\vec{x})\vec{x}}{r^{5}}\right).\label{eq:FGM}
\end{eqnarray}
For detailed discussions of the analogies between electrodynamics
and the linearized dynamics of GR, please consults \cite{Thorne1988,Ciufolini1995,Maartens1998}. Here,
such a separation of the relativistic perturbations will help us in solving the PN circular orbit. Along an unperturbed
Keplerian circular orbit, one
notices that the PN GE force becomes a centrifugal one with constant
magnitude
\[
\vec{f}_{GE}=\frac{M}{r^{3}}(\frac{4M}{r}-v^{2})\vec{x},
\]
and the projection of the GM force along the radial direction is also
a constant
\[
f_{GM}^{r}=\vec{f}_{GM}\cdot\frac{\vec{x}}{r}=2\frac{\vec{J}\cdot\vec{L}}{r^{4}},
\]
where $\vec{L}$ is the angular momentum of the orbiting satellite or proof mass.
Thus $\vec{f}_{GE}$ together with $\vec{f}_{GM}^{r}$ will only modify the total centripetal
force and therefore give rise to a relativistic
correction to the mean angular frequency $\omega$. The residual part
of the PN perturbations is a
periodic force $\vec{f}_{GM}^{\perp}$ that transverse to the orbit
plane, which will drive the orbit plane to precess about the direction of $\vec{J}$
(the Lense-Thirring precession \cite{Lense1918}), see Fig. \ref{fig:orbit} for illustration.

According to the above discussions, for the ideal case of an orbit
with constant radius $a$, the PN perturbations of the orbital elements
can be solved analytically up to the first order, therefore the nearly circular
orbit that satisfying the geodesic equation (\ref{eq:geodesic_tau})
can be worked out to the PN level as
\begin{eqnarray}
x^{1} & = & a\cos\Psi\cos\left(\frac{2J\tau}{a^{3}}\right)-a\cos i\sin\Psi\sin\left(\frac{2J\tau}{a^{3}}\right),
\label{eq:x1}\\
x^{2} & = & a\cos i\sin\Psi\cos\left(\frac{2J\tau}{a^{3}}\right)+a\cos\Psi\sin\left(\frac{2J\tau}{a^{3}}\right),
\label{eq:x2}\\
x^{3} & = & a\sin i\sin\Psi,\label{eq:x3}
\end{eqnarray}
where $i$ denotes the inclination, $\Psi=\omega\tau$ is the true
anomaly and the initial longitude of ascending node $\Omega(0)$
is set to be zero. For clarity, the time parameter
is replaced back to the proper time $\tau$ measured along the
above orbit, that from the differential line element $d\tau^{2}=-g_{\mu\nu}dx^{\mu}dx^{\nu}$
along the orbits the ratio $\frac{dt}{d\tau}$ can be solved
\begin{eqnarray}
\frac{dt}{d\tau} & = & 1+\frac{a^{2}\omega^{2}}{2}+\frac{M}{a}-\frac{a^{4}\omega^{4}}{8}+\frac{3Ma\omega^{2}}{2}+\frac{M^{2}}{2a^{2}}.\label{eq:dtoverdtau}
\end{eqnarray}
The PN modified angular frequency with respect
to $\tau$ can be worked out as
\begin{equation}
\omega=\sqrt{\frac{M}{a^{3}}}-\frac{3J\cos i}{M}.\label{eq:omega}
\end{equation}
Due to the frame-dragging effect \cite{Ciufolini1995}, the orbit plane precesses extremely slowly with rate $\dot{\Omega}(\tau)=\frac{2J}{a^{3}}$ about the direction of $\vec{J}$.
For a general low or medium Earth orbit it will take about $\frac{c^2\pi a^{3}}{G J}\sim10^{7}\ yrs$
to finish one period.

We will work with the orbit given in Eq. (\ref{eq:x1})-(\ref{eq:omega}) in this work. Perturbations
from small eccentricities and the corresponding PN effects are
left for future studies. The tidal matrix $K_{\mu\nu}$ defined in Eq.
(\ref{eq:Kdefinition}) along the this
orbit can be worked out under the geocentric coordinates system as
\begin{eqnarray}
(K^{N})_{\mu\nu}=\frac{M}{a^{3}}\left(\begin{array}{cccc}
0 & 0 & 0 & 0\\
\\
0 & -\frac{1}{2}(3\cos2\Psi+1) & -\frac{3}{2}\cos i\sin2\Psi & -\frac{3}{2}\sin
i\sin2\Psi\\
\\
0 & -\frac{3}{2}\cos i\sin2\Psi & \frac{1}{4}(6\cos2\Psi\cos^{2}i-3\cos2i+1) &
-3\cos i\sin i\sin^{2}\Psi\\
\\
0 & -\frac{3}{2}\sin i\sin2\Psi & -3\cos i\sin i\sin^{2}\Psi &
\frac{1}{4}(6\cos2\Psi\sin^{2}i+3\cos2i+1)
\end{array}\right),\nonumber\\\label{eq:KN}
\end{eqnarray}
\begin{eqnarray}
 &&(K^{GE})_{\mu\nu}=\nonumber\\
&& \frac{M}{a^{3}}\left(\begin{array}{cccc}
a^{2}\omega^{2} & a\omega\sin\Psi & -a\omega\cos i\cos\Psi & -a\omega\sin i\cos\Psi\\
\\
a\omega\sin\Psi & \frac{M}{2a}(1+3\cos2\Psi) & (\frac{3M}{2a}-2a^{2}\omega^{2})\cos i\sin2\Psi & (\frac{3M}{2a}-2a^{2}\omega^{2})\sin i\sin2\Psi\\
 & -a^{2}\omega^{2}(1+2\cos2\Psi)\\
\\
 &  & a^{2}\omega^{2}(2\cos2\Psi\cos^{2}i) & a^{2}\omega^{2}\sin2i(\cos2\Psi-2)\\
-a\omega\cos i\cos\Psi & (\frac{3M}{2a}-2a^{2}\omega^{2})\cos i\sin2\Psi & -a^{2}\omega^{2}(2\cos2i-1) & +\frac{3M\sin2i\sin^{2}\Psi}{2a}\\
 &  & +\frac{M(-6\cos2\Psi\cos^{2}i+3\cos2i-1)}{4a}\\
\\
 &  & a^{2}\omega^{2}\sin2i(\cos2\Psi-2) & a^{2}\omega^{2}(2\cos2\Psi\sin^{2}i)\\
-a\omega\sin i\cos\Psi & (\frac{3M}{2a}-2a^{2}\omega^{2})\sin i\sin2\Psi & +\frac{3M\sin2i\sin^{2}\Psi}{2a} & +a^{2}\omega^{2}(2\cos2i+1)\\
 &  &  & -\frac{M(6\cos2\Psi\sin^{2}i+3\cos2i+1)}{4a}
\end{array}\right),\nonumber \\
\label{eq:KGE}
\end{eqnarray}
\begin{eqnarray}
&& (K^{GM})_{\mu\nu}=\nonumber\\
&&\frac{3J\omega}{a^{3}}\left(\begin{array}{cccc}
0 & 0 & 0 & 0\\
\\
0 & 2\cos i\cos^{2}\Psi & \frac{1}{2}(3\cos2i-1)\sin2\Psi & 3\sin2i\sin\Psi\cos\Psi\\
\\
0 & \frac{1}{2}(3\cos2i-1)\sin2\Psi & \cos i(5\cos2i-3)\sin^{2}\Psi & \frac{1}{4}(10\sin3i\sin^{2}\Psi)\\
 &  &  & +\frac{1}{4}(3\cos2\Psi\sin i+\sin i)\\
\\
0 & 3\sin2i\sin\Psi\cos\Psi & \frac{1}{4}(10\sin3i\sin^{2}\Psi) & -\frac{1}{4}\cos i(20\cos2\Psi\sin^{2}i+3)\\
 &  & +\frac{1}{4}(3\cos2\Psi\sin i+\sin i) & -\frac{5}{4}\cos3i
\end{array}\right),\nonumber \\
\label{eq:KGM}
\end{eqnarray}
and within mission life-time $T$ much smaller
compared with the period of the Lense-Thirring precession, that  $\frac{T}{a}\ll\frac{\pi a^{2}}{J}$,
we also have
\begin{eqnarray}
(K^{LT})_{\mu\nu}=\frac{3JM\Psi}{a^{6}\omega}\left(\begin{array}{cccc}
0 & 0 & 0 & 0\\
\\
0 & 2\cos i\sin2\Psi & -\frac{1}{2}(\cos2i+3)\cos2\Psi-\sin^{2}i &
\sin2i\sin^{2}\Psi\\
\\
0 &  -\frac{1}{2}(\cos2i+3)\cos2\Psi-\sin^{2}i & -2\cos i\sin2\Psi & -\sin
i\sin2\Psi\\
\\
0 & \sin2i\sin^{2}\Psi & -\sin i\sin2\Psi&0
\end{array}\right).\nonumber\\
\label{eq:KS}
\end{eqnarray}
Here, $K_{\mu\nu}$ is decomposed
into four parts, the Newtonian part %(of order $\frac{1}{a^{2}}\mathcal{O}(\epsilon^{2})$)
$(K^{N})_{\mu\nu}$, the PN GE part $(K^{GE})_{\mu\nu}$,
the GM part $(K^{GM})_{\mu\nu}$ and the secular part $(K^{LT})_{\mu\nu}.$
The secular terms in $(K^{LT})_{\mu\nu}$ are in fact periodic ones with periods $\sim \frac{\pi a^3}{J}$,
which are produced by the modulation of the Newtonian tidal tensor
due to the Lense-Thirring precession of the orbit with respect to the geocentric coordinates system.
For mission life time $T$ about a few years with total orbital cycles $\frac{\Psi}{2\pi}\sim10^{4}$,
the secular tensor is of the PN level $|(K^{LT})_{\mu\nu}|\sim\frac{JMT}{a^{6}}\sim\frac{1}{a^{2}}\Psi\mathcal{O}(\epsilon^{4})$.

\section{Local Frame Along Post-Newtonian Orbit\label{sec:Local Frame}}

\begin{figure}
\includegraphics[scale=0.55]{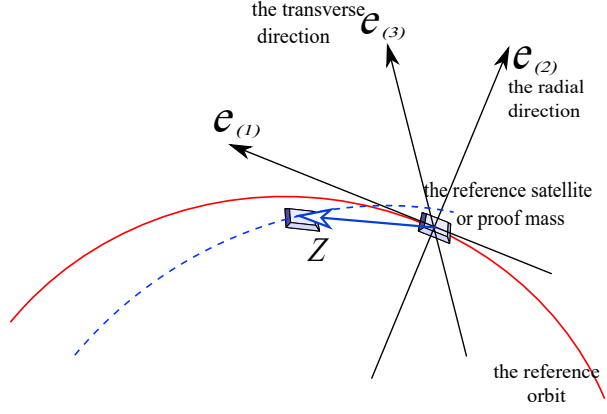}
\caption{The local frame attached to the reference satellite or proof mass.}
\label{fig:LF}
\end{figure}
In this section, we work out the tetrad
$e_{(a)}^{\ \ \ \mu}$ attached to the satellite or proof mass that
following the orbit given in Eq. (\ref{eq:x1})-(\ref{eq:x3}). The relativistic
corrections up to the PN level in the tetrad %, that $e_{(a)}^{\ \ \ \mu}=I+\mathcal{O}(\epsilon^2)$,
need to be included.

We first set $e_{(0)}^{\ \ \ \mu}=\tau^{\mu}=\frac{dx^{\mu}}{d\tau}$,
which is the 4-velocity of the reference  satellite or proof mass.
Along the orbit given in Eq. (\ref{eq:x1})-(\ref{eq:x3}), we have
\begin{equation}
e_{(0)}^{\ \ \ \mu}=\tau^{\mu}=\left(\begin{array}{c}
1+\frac{a^{2}\omega^{2}}{2}+\frac{M}{a}\\
-a\omega\sin\Psi-\frac{2J\cos i(\sin\Psi+\Psi\cos\Psi)}{a^{2}}\\
a\omega\cos i\cos\Psi+\frac{2J(\cos\Psi-\Psi\sin\Psi)}{a^{2}}\\
a\omega\sin i\cos\Psi
\end{array}\right).\label{eq:e0}
\end{equation}
Second, we set the orthonormal spacial triad $\vec{e}_{(i)}$ as follows,
that $\vec{e}_{(1)}$ is parallel to the direction of the instance 3-velocity
$\vec{v}=\frac{d\vec{x}}{dt}$ of the reference satellite or proof
mass, $\vec{e}_{(2)}$ is along the radial direction $\vec{x}$ and
$\vec{e}_{(3)}=\vec{e}_{(1)}\times\vec{e}_{(2)}$, see Fig. \ref{fig:LF}
for illustration. Last but not least, since the local frame is moving
along the world line of the reference satellite or proof mass,
one needs to perform the boost Lorentz transformations
of the triad with respect to the 4-velocity $\tau^{\mu}$. Within
the short time limits that $\frac{\tau}{a}\ll\frac{\pi a^{2}}{J}$
we then have

\begin{equation}
e_{(1)}{}^{\mu}=\left(\begin{array}{c}
\omega(a-M)+\frac{2J\cos i}{a^{2}}\\
-\left(1-\frac{M}{a}+\frac{a^{2}\omega^{2}}{2}\right)\sin\Psi-\frac{2J\Psi\cos i\cos\Psi}{a^{3}\omega}\\
\left(1-\frac{M}{a}+\frac{a^{2}\omega^{2}}{2}\right)\cos i\cos\Psi+\frac{2J\sin^{2}i\cos\Psi-2J\Psi\sin\Psi}{a^{3}\omega}\\
\left(1-\frac{M}{a}+\frac{a^{2}\omega^{2}}{2}\right)\sin i\cos\Psi-\frac{2J\cos i\sin i\cos\Psi}{a^{3}\omega}
\end{array}\right),\label{eq:e1}
\end{equation}
\begin{equation}
e_{(2)}{}^{\mu}=\left(\begin{array}{c}
0\\
\left(1-\frac{M}{a}\right)\cos\Psi-\frac{2J\Psi\cos i\sin\Psi}{a^3\omega}\\
\left(1-\frac{M}{a}\right)\cos i\sin\Psi+\frac{2J\Psi\cos\Psi}{a^3\omega}\\
\left(1-\frac{M}{a}\right)\sin i\sin\Psi
\end{array}\right),\label{eq:e2}
\end{equation}
\begin{equation}
e_{(3)}{}^{\mu}=\left(\begin{array}{c}
0\\
\frac{J\sin i(\sin2\Psi-2\Psi)}{a^{3}\omega}\\
\left(1-\frac{M}{a}-\frac{2J\cos i\cos^{2}\Psi}{a^{3}\omega}\right)\sin i\\
\left(-1+\frac{M}{a}\right)\cos i-\frac{2J\sin^{2}i\cos^{2}\Psi}{a^{3}\omega}
\end{array}\right).\label{eq:e3}
\end{equation}
Therefore, the tetrad matrix $e_{(a)}^{\ \ \ \mu}$, which can be viewed
as the transformation matrix from the local frame to the geocentric
coordinates system, and its inverse{\small{} }$e^{(a)}_{\ \ \ \mu}$ can be worked out as
{\small{}
\begin{eqnarray}
 e_{(a)}{}^{\mu}=\left(\begin{array}{cccc}
1+\frac{a^{2}\omega^{2}}{2}+\frac{M}{a} & -a\omega\sin\Psi & a\omega\cos i\cos\Psi & a\omega\sin i\cos\Psi\\
 & -\frac{2J\cos i(\Psi\cos\Psi+\sin\Psi)}{a^{2}} & +\frac{2J(\cos\Psi-\Psi\sin\Psi)}{a^{2}}\\
\\
(a-M)\omega & -(1-\frac{M}{a}+\frac{a^{2}\omega^{2}}{2})\sin\Psi & (1-\frac{M}{a}+\frac{a^{2}\omega^{2}}{2})\cos i\cos\Psi & (1-\frac{M}{a}+\frac{a^{2}\omega^{2}}{2})\sin i\cos\Psi\\
 +\frac{2J\cos i}{a^{2}} & -\frac{2J\Psi\cos i\cos\Psi}{a^{3}\omega} & +\frac{2J\sin^{2}i\cos\Psi-2J\Psi\sin\Psi}{a^{3}\omega} & -\frac{2J\cos i\sin i\cos\Psi}{a^{3}\omega}\\
\\
0 & (1-\frac{M}{a})\cos\Psi & (1-\frac{M}{a})\cos i\sin\Psi  & (1-\frac{M}{a})\sin i\sin\Psi\\
 & -\frac{2J\Psi\cos i\sin\Psi}{a^3\omega}& +\frac{2J\Psi\cos\Psi}{a^3\omega} \\
\\
0 & \frac{J\sin i(\sin 2\Psi-2\Psi)}{a^{3}\omega} & (1-\frac{M}{a})\sin i & (-1+\frac{M}{a})\cos i\\
 &  & -\frac{2J\sin i\cos i\cos^{2}\Psi}{a^{3}\omega} & -\frac{2J\cos^{2}\Psi\sin^{2}i}{a^{3}\omega}
\end{array}\right),\nonumber \\
\label{eq:e}
\end{eqnarray}
\begin{eqnarray}
 e^{(a)}{}_{\mu}=
 \left(\begin{array}{cccc}
1+\frac{a^{2}\omega^{2}}{2}-\frac{M}{a} & -a\omega-\frac{2J\cos i}{a^{2}} & 0 & 0\\
\\
(a-M)\omega\sin\Psi & -(1+\frac{\omega^{2}a^{2}}{2}+\frac{M}{a})\sin\Psi & (1+\frac{M}{a})\cos\Psi & -\frac{J\sin i(2\Psi-\sin 2\Psi)}{a^{3}\omega}\\
+\frac{2J\cos i(\Psi\cos\Psi+\sin\Psi)}{a^{2}} & -\frac{2J\Psi\cos i\cos\Psi}{a^{3}\omega} & -\frac{2J\Psi\cos i\sin\Psi}{a^{3}}\\
\\
(M-a)\omega\cos i\cos\Psi & (1+\frac{\omega^{2}a^{2}}{2}+\frac{M}{a})\cos i\cos\Psi & (1+\frac{M}{a})\cos i\sin\Psi & (1+\frac{M}{a})\sin i\\
+\frac{2J(\Psi\sin\Psi-\cos\Psi)}{a^{2}} & +\frac{2J\sin^{2}i\cos\Psi-2J\Psi\sin\Psi}{a^{3}\omega} & +\frac{2J\Psi\cos\Psi}{a^{3}\omega} & -\frac{2J\cos i\sin i\cos^{2}\Psi}{a^{3}\omega}\\
\\
(M-a)\omega\sin i\cos\Psi & (1+\frac{\omega^{2}a^{2}}{2}+\frac{M}{a})\sin i\cos\Psi & (1+\frac{M}{a})\sin i\sin\Psi & -(1+\frac{M}{a})\cos i\\
 & -\frac{2J\cos i\sin i\cos\Psi}{a^{3}\omega} &  & -\frac{2J\sin^{2}i\cos^{2}\Psi}{a^{3}\omega}
\end{array}\right).\nonumber \\
\label{eq:ie}
\end{eqnarray}

Under such local frame, the tidal matrix $K_{(a)(b)}$ in Eq. (\ref{eq:Klocal})
can be derived. As expected, up to the PN level the components
$K_{(0)(a)}=0$, and the Newtonian spatial part
reads
\begin{equation}
(K^{N})_{(i)(j)}=\left(\begin{array}{ccc}
\frac{M}{a^{3}} & 0 & 0\\
0 & -\frac{2M}{a^{3}} & 0\\
0 & 0 & \frac{M}{a^{3}}
\end{array}\right)\label{eq:KabN}
\end{equation}
which agrees exactly with the Newtonian tidal tensor $\partial_{i}\partial_{j}U$
evaluated in the LVLH frame along circular orbits. The PN parts turns our to be
\begin{equation}
(K^{GE})_{(i)(j)}=\left(\begin{array}{ccc}
-\frac{3M^{2}}{a^{4}} & 0 & 0\\
0 & \frac{3M(2M-a^{3}\omega^{2})}{a^{4}} & 0\\
0 & 0 & -\frac{3M(M-a^{3}\omega^{2})}{a^{4}}
\end{array}\right),\label{eq:KabGE}
\end{equation}
\begin{equation}
(K^{GM})_{(i)(j)}=\left(\begin{array}{ccc}
0 & 0 & \frac{3J\omega\sin i\cos\Psi}{a^{3}}\\
0 & \frac{6J\omega\cos i}{a^{3}} & -\frac{9J\omega\sin i\sin\Psi}{a^{3}}\\
\frac{3J\omega\sin i\cos\Psi}{a^{3}} & -\frac{9J\omega\sin i\sin\Psi}{a^{3}} & -\frac{6J\omega\cos i}{a^{3}}
\end{array}\right).\label{eq:KabGM}
\end{equation}
 }{\small \par}

\section{Post-Newtonian Extension of Hill-Clohessy-Wiltshire Equations\label{sec:PNHCW}}

Now, with all the results that gathered in previous sections, we substitute
the orbit given in Eq. (\ref{eq:x1})-(\ref{eq:omega}), the Christoffel
symbols in Eq. (\ref{eq:G1})-(\ref{eq:G6}), the tetrad matrices
in Eq. (\ref{eq:e}) and (\ref{eq:ie}), and the tidal matrices in
Eq. (\ref{eq:KabN}) - (\ref{eq:KabGM}) into the geodesic deviation
equation (\ref{eq:localdev}). With straightforward but tedious algebraic
manipulations and leaving out all the terms beyond $\frac{|Z|}{a^{2}}\mathcal{O}(\epsilon^{4})$
and $\frac{|Z|}{a^{2}}\Psi\mathcal{O}(\epsilon^{4})$, the geodesic
deviation equation (\ref{eq:localdev}) under the local frame can be
cast into an elegant form. Recovering the SI units, the final form of the
Post-Newtonian extension of the Hill-Clohessy-Wiltshire equations along a relativistic
nearly circular orbit turns out to be
\begin{eqnarray}
\left(\begin{array}{c}
\ddot{Z}^{(1)}(\tau)\\
\ddot{Z}^{(2)}(\tau)\\
\ddot{Z}^{(3)}(\tau)
\end{array}\right)+\underbrace{\left(\begin{array}{ccc}
0 & 0 & 0\\
0 & -3\omega^{2} & 0\\
0 & 0 & \omega^{2}
\end{array}\right)}_{Newtonian\ gradient}\left(\begin{array}{c}
Z^{(1)}(\tau)\\
Z^{(2)}(\tau)\\
Z^{(3)}(\tau)
\end{array}\right)+\underbrace{\left(\begin{array}{ccc}
0 & 2\omega & 0\\
-2\omega & 0 & 0\\
0 & 0 & 0
\end{array}\right)}_{Coriolis}\ \left(\begin{array}{c}
\dot{Z}^{(1)}(\tau)\\
\dot{Z}^{(2)}(\tau)\\
\dot{Z}^{(3)}(\tau)
\end{array}\right)\nonumber \\
+\underbrace{\left(\begin{array}{ccc}
0 & 0 & \frac{4GJ\omega\sin i\cos(\tau\omega)}{c^{2}a^{3}}\\
0 & \frac{6a^{2}\omega^{4}}{c^{2}}-\frac{12GJ\omega\cos i}{c^{2}a^{3}} & -\frac{10GJ\omega\sin i\sin(\tau\omega)}{c^{2}a^{3}}\\
0 & -\frac{12GJ\omega\sin i\sin(\tau\omega)}{c^{2}a^{3}} & 0
\end{array}\right)}_{PN\ gradient}\left(\begin{array}{c}
Z^{(1)}(\tau)\\
Z^{(2)}(\tau)\\
Z^{(3)}(\tau)
\end{array}\right)\nonumber \\
+\underbrace{\left(\begin{array}{ccc}
0 & \frac{6GJ\cos i}{c^{2}a^{3}}-\frac{3a^{2}\omega^{3}}{c^{2}} & \frac{4GJ\sin i\sin(\tau\omega)}{c^{2}a^{3}}\\
\frac{3a^{2}\omega^{3}}{c^{2}}-\frac{6GJ\cos i}{c^{2}a^{3}} & 0 & -\frac{2GJ\sin i\cos(\tau\omega)}{c^{2}a^{3}}\\
-\frac{4GJ\sin i\sin(\tau\omega)}{c^{2}a^{3}} & \frac{2GJ\sin i\cos(\tau\omega)}{c^{2}a^{3}} & 0
\end{array}\right)}_{PN\ corrections\ to\ Coriolis}\left(\begin{array}{c}
\dot{Z}^{(1)}(\tau)\\
\dot{Z}^{(2)}(\tau)\\
\dot{Z}^{(3)}(\tau)
\end{array}\right) & = & 0,\nonumber \\
\label{eq:PNCW}
\end{eqnarray}
with the time component of the geodesic deviation equation (\ref{eq:localdev})
has the trivial form as expected
\[
\ddot{Z}^{0}(\tau)=0.
\]
One notices that if all the PN corrections, that the second and the last lines of the above equation, are left out,
the classical HCW equations, that the first line in the above equation,
can be recovered. The second line in the above equation results from the combination of
the PN tidal forces from spacetime curvature and the PN corrections to inertia forces caused by the geodetic
\cite{deSitter1916} and Schiff \cite{Schiff1960} precessions of the local inertia frame along the orbit.
The last line comes from the PN corrections to the Coriolis forces.
Eq. (\ref{eq:PNCW}) is the key result of this work,
which may be taken as the foundation and starting point
of the studies of relativistic dynamics of relative orbit motions.

As a demonstration of possible applications of the extended equations,
we end up this section with the PN corrections to the general periodic solutions
of the classical HCW equations
\[
\left(\begin{array}{c}
\ddot{Z}^{(1)}(\tau)\\
\ddot{Z}^{(2)}(\tau)\\
\ddot{Z}^{(3)}(\tau)
\end{array}\right)+\left(\begin{array}{ccc}
0 & 0 & 0\\
0 & -3\omega^{2} & 0\\
0 & 0 & \omega^{2}
\end{array}\right)\left(\begin{array}{c}
Z^{(1)}(\tau)\\
Z^{(2)}(\tau)\\
Z^{(3)}(\tau)
\end{array}\right)+\left(\begin{array}{ccc}
0 & 2\omega & 0\\
-2\omega & 0 & 0\\
0 & 0 & 0
\end{array}\right)\left(\begin{array}{c}
\dot{Z}^{(1)}(\tau)\\
\dot{Z}^{(2)}(\tau)\\
\dot{Z}^{(3)}(\tau)
\end{array}\right)=0.
\]
With the initial values $\{\dot{Z}^{(i)}(0)=\dot{Z}_{0}^{(i)},\ Z^{(i)}(0)=Z_{0}^{(i)}\}$,
the general solutions have the form
\begin{eqnarray*}
Z^{(1)}(\tau)&= & Z_{0}^{(1)}+6Z_{0}^{(2)}\left(\sin(\omega\tau)-\omega\tau\right)+\frac{4\dot{Z}_{0}^{(1)}\sin(\omega\tau)}{\omega}\nonumber\\
 && -3\dot{Z}_{0}^{(1)}\tau+\frac{2\dot{Z}_{0}^{(2)}\left(\cos(\omega\tau)-1\right)}{\omega},\\
Z^{(2)}(\tau)&= & Z_{0}^{(2)}(4-3\cos(\omega\tau))+\frac{2\dot{Z}_{0}^{(1)}(1-\cos(\omega\tau))}{\omega}+\frac{\dot{Z}_{0}^{(2)}\sin(\omega\tau)}{\omega},\\
Z^{(3)}(\tau)&= & Z_{0}^{(3)}\cos(\omega\tau)+\frac{\dot{Z}_{0}^{(3)}\sin(\omega\tau)}{\omega}.
\end{eqnarray*}
To remove the drift terms we set $\dot{Z}_{0}^{(1)}=0$ and $Z_{0}^{(2)}=0$,
and the general periodic solutions $Z_p^{(i)}(\tau)$ of the HCW equations read
\begin{eqnarray}
Z_{P}^{(1)}(\tau)&= & Z_{0}^{(1)}+\frac{2\dot{Z}_{0}^{(2)}(\cos(\omega\tau)-1)}{\omega},\label{eq:Zp1}\\
Z_{P}^{(2)}(\tau)&= & \frac{\dot{Z}_{0}^{(2)}\sin(\omega\tau)}{\omega},\label{eq:Zp2}\\
Z_{P}^{(3)}(\tau)&= & Z_{0}^{(3)}\cos(\omega\tau)+\frac{\dot{Z}_{0}^{(3)}\sin(\omega\tau)}{\omega}.\label{eq:Zp3}
\end{eqnarray}
These solutions had already found many applications in the literature. For future SST missions and
missions with high-precision optical gradiometers (as demonstrated in LPF), the PN corrections
to the above periodic solutions may be important to the measurements, especially to those in the along track direction.
Let us assume the PN solutions
to be $Z_{P}^{(i)}(\tau)+\delta^{(i)}(\tau)$ with $\delta^{(i)}(\tau)\sim|Z_{P}|\mathcal{O}(\epsilon^{2})$
the PN corrections to the periodic solution given in Eq. (\ref{eq:Zp1})-(\ref{eq:Zp3}).
Substitute $Z_{P}^{(i)}(\tau)+\delta^{(i)}(\tau)$ into Eq. (\ref{eq:PNCW})
and leaving out terms beyond $\frac{|Z_{P}|}{a^{2}}\mathcal{O}(\epsilon^{4})$,
we have
\begin{eqnarray}
\ddot{\delta}^{(1)}(\tau)&= & -2\omega\dot{\delta}^{(2)}(\tau)+\frac{3a^{2}\dot{Z}_{0}^{(2)}\omega^{3}}{c^{2}}\cos(\tau\omega)+\frac{4GZ_{0}^{(3)}J\omega\sin i\sin^{2}(\omega\tau)}{c^{2}a^{3}}\nonumber\\
 && -\frac{4GZ_{0}^{(3)}J\omega\sin i\cos^{2}(\omega\tau)+6G\dot{Z}_{0}^{(2)}J\cos i\cos(\omega\tau)}{c^{2}a^{3}}\nonumber\\
 && -\frac{4G\dot{Z}_{0}^{(3)}J\sin i\sin(2\omega\tau)}{c^{2}a^{3}},\label{eq:PNp1}\\
\ddot{\delta}^{(2)}(\tau)&= & 3\omega^{2}\delta^{(2)}(\tau)+2\omega\dot{\delta}^{(1)}(\tau)+\frac{4GZ_{0}^{(3)}J\omega\sin i\sin(2\omega\tau)}{c^{2}a^{3}}\nonumber\\
 && +\frac{10G\dot{Z}_{0}^{(3)}J\sin i\sin^{2}(\omega\tau)}{c^{2}a^{3}}+\frac{2G\dot{Z}_{0}^{(3)}J\sin i\cos^{2}(\omega\tau)}{c^{2}a^{3}},
 \label{eq:PNp2}\\
\ddot{\delta}^{(3)}(\tau)&= & -\omega^{2}\delta^{(3)}(\tau)-\frac{GJ\dot{Z}_{0}^{(2)}\sin i(3\cos(2\omega\tau)-1)}{c^{2}a^{3}}\label{eq:PNp3}.
\end{eqnarray}
With the initial conditions $\{\dot{\delta}_{0}^{(i)}=\delta_{0}^{(i)}=0\}$,
the general solutions can be worked out as
\begin{eqnarray}
\delta^{(1)}(\tau)&= & \frac{3a^{2}\dot{Z}_{0}^{(2)}\omega}{c^{2}}(2\tau\omega\sin(\omega\tau)+3\cos(\omega\tau)-3)\nonumber\\
 & &-\frac{6G\dot{Z}_{0}^{(2)}J\cos i(2\tau\omega\sin(\omega\tau)+3\cos(\omega\tau)-3)}{c^{2}a^{3}\omega^{2}}\nonumber\\
 & &+\frac{GZ_{0}^{(3)}J\sin i(\cos(2\omega\tau)-1)}{c^{2}a^{3}\omega}\nonumber\\
 & &+\frac{G\dot{Z}_{0}^{(3)}J\sin i(4\sin(\omega\tau)+\sin(2\omega\tau)-6\tau\omega)}{c^{2}a^{3}\omega^{2}},\label{eq:delta1}\\
\delta^{(2)}(\tau)&= & -\frac{3a^{2}\dot{Z}_{0}^{(2)}\omega}{c^{2}}(\tau\omega\cos(\omega\tau)-\sin(\omega\tau))\nonumber\\
 & &-\frac{6G\dot{Z}_{0}^{(2)}J\cos i(\sin(\omega\tau)-\tau\omega\cos(\omega\tau))}{c^{2}a^{3}\omega^{2}}\nonumber\\
 & &-\frac{2G\dot{Z}_{0}^{(3)}J\sin i(\cos(\omega\tau)-1)}{c^{2}a^{3}\omega^{2}},\label{eq:delta2}\\
\delta^{(3)}(\tau)&= & -\frac{4G\dot{Z}_{0}^{(2)}J\sin i\sin^{2}\left(\frac{\omega\tau}{2}\right)\cos(\omega\tau)}{c^{2}a^{3}\omega^{2}}\label{eq:delta3}.
\end{eqnarray}
There exist a drift term in $\delta^{(1)}(\tau)$ in the last line of Eq. (\ref{eq:delta1}),
which may be further removed by assuming $\dot{Z}_{0}^{(3)}=0$.
The solutions also contain oscillating terms with growing magnitudes in $\delta^{(1)}(\tau)$ and $\delta^{(2)}(\tau)$ within
the $e_{(1)}^{\ \ \ \mu}-e_{(2)}^{\ \ \ \mu}$ plane (the orbital plane), which remain true only when the conditions
$|\delta^{(i)}(\tau)|\ll |Z_{p}|$ are satisfied. For applications, even after $\frac{\tau \omega}{2\pi}\sim 10^4$ orbital cycles,
one has the magnitudes estimation $|\delta^{(i)}(\tau)| \sim |Z_p|\tau\omega\mathcal{O}(\epsilon^2)\leq 10^{-6}\times|Z_{p}|$, and still these solutions can be taken as good approximations.

\section{Conclusions\label{sec:Conclusion}}
This work can be viewed as the first-step study of the
relativistic dynamics of relative orbit motions. A systematic approach to the linearized theory through the
geodesic deviation equation in GR is introduced. When the centered source is modeled as an ideal uniform and rotating spherical body, the relativistic equations, that Eq. (\ref{eq:PNCW}), determining the relative motions with
respect to a relativistic circular orbit are derived up to the PN level. These equations
are the PN extensions of the classical HCW equations, which may be taken as the starting point for future studies
of relativistic  relative orbit motions.
The PN corrections given in Eq. (\ref{eq:delta1})-(\ref{eq:delta3}) to the periodic solutions of the HCW
equations, especially the growing oscillating terms, show that
general relativistic effects may be important to both inter-satellites ranges measurements and space-borne gradient measurements. While, for practical applications of the generalized HCW equations to such problems, one needs to work with the much more complicated metric given in Eq. (\ref{eq:full_metric}) to enclose perturbations from certain geopotential multipoles, choose a more
accurate reference orbit and deal with the related errors in a more sophisticated way.
Such topics, as natural subsequent works, will be left for future studies.

\section*{Acknowledgments}
Supports from National Space Science Center, Chinese
Academy of Sciences (No. XDA04077700), National Natural
Science Foundation of China (No. 11171329
and No. 41404019)  and  Central Universities
Funds (No. 2014G3262010 and No. 310826161010) are acknowledged. We are
grateful to Yun Kau Lau for initiating the study of the
problem and encouraging us to do this work. The author Peng Xu is also
grateful to Shing-Tung Yau for his continuous
support at the Morningside Center
of Mathematics, Chinese Academy of Sciences.
Valuable discussions with Dr. Wenlin Tang is acknowledged.

\end{document}